\documentclass[final]{my}
\usepackage{times}
\begin{document}
\lineskip 3pt \normallineskip 3pt
\catcode`@=11
\newtheorem{@MyTheorem}{Theorem}
\newtheorem{@MyProposition}[@MyTheorem]{Proposition}
\newtheorem{@MyLemma}[@MyTheorem]{Lemma}
\newtheorem{@MyDefinition}[@MyTheorem]{Definition}
\newtheorem{@MyClaim}[@MyTheorem]{Claim}
\newtheorem{@MyRemark}[@MyTheorem]{Remark}
\newtheorem{@MyConjecture}[@MyTheorem]{Conjecture}
\def\eightpoint{}
\def\be#1{\begin{equation}\label{eq@#1}}
\def\FIGUREWITHTEX#1#2#3#4#5#6#7#8{
\begingroup
\footnotesize
\def\epsfsize##1##2{#4##1}
\unitlength=#4bp
\newdimen\figh
\figh=#3\unitlength
\setbox0=\hbox to \textwidth{\epsfbox{#1}\hfill}
\figh=\ht0
\providecommand{\figinput}[1]{\input {##1}}
\setbox1=\hbox to\textwidth{\input {##2}\hfill}
\begin{figure}
\setbox2=\vbox{\box0\hskip-\wd0\hskip#5truecm
\vskip#6truecm\box1\hskip-#5truecm\vskip-#6truecm}
\vbox{\hskip#7truecm\box2\hfill}
\caption{#8}\label{figure@#1}
\end{figure}
\endgroup
}
\def\fig(#1){Fig.~\ref{figure@#1}}
\def\REMARK{\LIKEREMARK{Remark}}
\def\text#1{\leavevmode\hbox{#1}}
\def\equ(#1){(\ref{eq@#1})}
\catcode`@=12
\newcommand\operatorname[1]{{\rm #1}}
\newcommand\tr{\operatorname{tr}}
\newcommand\supp{\operatorname{supp}}
\newcommand\dist{\operatorname{dist}}
\newcommand\Id{\operatorname{Id}}
\newcommand\diam{\operatorname{diam}}
\newcommand\Jac{\operatorname{Jac}}
\newcommand\esssup{\operatornamewithlimits{ess\phantom{:}sup}}
\newcommand\essinf{\operatornamewithlimits{ess\phantom{:}inf}}
\newcommand\spt{\operatorname{spt}}
\newcommand\dimp{{\rm dim_p}}
\newcommand\dimH{{\rm dim_H}}
\newcommand\real{{\bf R}}
\def\m(#1){\mu(B(#1))}
\def\L{{\text{L}}}
\def\R{{\text{R}}}
\def\ie{{\it i.e.}}
\def\eg{{\it e.g.}}
\def\por{{\rm por}}
\let\epsilon=\varepsilon
\def\LIKEREMARK#1{\medskip\noindent{\bf #1\ }}
\title{Porosities and dimensions of measures satisfying the doubling condition}

\author{Jean-Pierre Eckmann}
\address{University of Geneva, Departments of Physics and Mathematics,\\
1211 Geneva 4, Switzerland\\E-mail: Jean-Pierre.Eckmann@physics.unige.ch}
\author{Esa J\"arvenp\"a\"a
 and Maarit J\"arvenp\"a\"a}
\address{  University of Jyv\"askyl\"a, Department of Mathematics,\\ P.O. Box 35,
  FIN-40351 Jyv\"askyl\"a, Finland\\
  E-mail: esaj@math.jyu.fi and
  amj@math.jyu.fi}
\maketitle
\abstracts{This paper explains the implications of a mathematical
theory (by the same authors) of 
holes in fractals and their relation to dimension for
measurements. The novelty of our
approach is to consider the fractal measure on a set rather than just the
support of that measure. This should take into account in a more
precise way the distribution of data points in measured sets, such as
the distribution of galaxies.
}

\section{Introduction}
The aim of this talk is to describe a mathematics paper\cite{EJJ} written
by the same authors to a community of scientists
not necessarily specialized in the theory of fractals. We hope that
the concepts developed here can help to understand some issues which
have occurred in connection with the debate about the dimension of the
galaxy distribution.

We want to explain the
intuitive motivations of our definitions and results in the context of
actual physical problems. These motivations will be underlined by
simple but relevant examples.

The observation and study of fractal sets, \ie, sets with
possibly non-integer dimension, is of course widespread. The present
work deals with two new aspects of fractal sets:

\begin{list}{}{}
\item{$\bullet$\ }Can the presence of voids be used to say something about
the dimension of a set?

\item{$\bullet$\ }How do voids which are almost empty account for the
dimension of a set?
\end{list}
We will give some answers in the case of sets which
are not too
wild, namely satisfying a certain doubling condition.

Our interest in the question of voids in fractals (porosity) was
raised by questions about the fractal dimension of galaxy
distributions. In view of the heated debate in this subject, see \eg
\cite{D,PMS},
it seems adequate to provide as many analytical tools as possible with
which the fractal dimension can be estimated. This paper provides a new
such tool, namely the {\em porosity of the measure on the set}. The
general idea is that sets with large voids must have small dimension.
These voids, \eg~in galaxy distributions, are taken as indicators of
small dimension in a sense we make precise below, and our theory
allows a systematic way to disregard occasional points (galaxies)
inside a (large) void. We present here a formalism which is tailored
for this situation, by presenting algorithms for measures rather than
for their supports.

\section{Measures and sets}

The main idea\cite{M1,S} describing the relation between the porosity and
the dimension goes about as follows, and we present some intuitive
examples which can guide the reader unfamiliar with this problem. Take
the well-known middle third Cantor set $C$. Clearly, if we consider
the voids in this set, every point in $C$ is close to a relatively
large void (namely to the middle third which has been taken out).
We will
give a more precise definition below.
Another way to view porosity, which is closer to the actual definition
is as follows: Suppose we have found a void of diameter $\rho$ inside
some minimal ball of radius $r'$ around a given point of the
fractal. Then the question is: How big a radius we have to take in order to 
see for the first time a bigger void than the one we have
already seen? See also \fig(lacundivp.ps) below.

It is clear that if we take out
more of the middle, we make the dimension of the set
smaller, since the dimension depends in a well-known fashion on the length
ratio of voids to non-voids, namely, if we remove an interval of length
$\frac{k-2}k$ from the middle, \ie, leave two intervals of length $\frac  1k$,
then the dimension is (in $\real$)
$\log 2/\log k$. For example for the middle third Cantor
set the dimension is $\log 2/\log 3$.
Thus, {\em large voids imply
small dimension}. However, the contrary is {\em not} true, as we shall
explain in Example 3 below.
One can construct a sequence of regular fractals, all of the
same dimension, but with porosity decreasing to 0.
Thus, {\em a set can have small dimension
without any porosity}. It is this aspect which is connected with the
controversy about the dimension of the galaxy distribution.

The requirement of obtaining information about
experimentally measurable objects leads us to consider measures, or
mass distributions, rather than sets.
This issue was addressed earlier\cite{ER} in the context of dimension
measurements. For example, one can compute the dimension of the
{\em support} of a set, \ie, of the complement of the open sets of zero
measure. But, as explained in\cite{ER}, the so-called correlation
dimension which is based on the {\em mass} in balls seems to be the natural
quantity for questions of experimental nature and is commonly used in
the Grassberger-Procaccia method\cite{GP}. 
Therefore, we shall study here the porosity of a {\em measure}, and
not only the porosity of the support of the measure
\cite{ER}.
We then show that large porosity implies a non-trivial upper bound on
the dimension (in fact on all multi-fractal dimensions
$D_q$, $q>1$). Finally, we explain how porosity is estimated for a given set
of experimental points.

\section{Porosities of measures}
Let $\mu$ be a probability measure on
$\real^n$. We define for $x\in \real^n$ and $r,\epsilon >0$:
\begin{eqnarray*}
  \por&(\mu,x,r,\varepsilon)\,
  =\,\sup\{{p}\ge 0~:  {\text{there is }z\in\real^n
   \text{ such that }} \\
&B(z,p r)\subset B(x,r)\text{ and }\mu(B(z,p
  r))\le\varepsilon\mu(B(x,r))\}~.
\end{eqnarray*}
In other words, we consider the ball $B(x,r)$ of radius $r$ centered at 
$x\in\real^n$ and observe the mass $\mu(B(x,r))$ contained in it.
We now look for
the largest ball of radius $p\cdot r$
(fully contained in $B(x,r)$) around a point
$z$ such that the mass of that ball does not exceed $\epsilon
$ times the mass $\mu(B(x,r))$ of $B(x,r)$.\footnote[1]{The conventional porosity asks for the largest ball
in $B(x,r)$ which does not contain {\em any} point of the set in question;
see also below.} 
See \fig(fig1p.ps).
One then defines
\begin{equation}\label{eq@pmx}
\por(\mu,x)\,=\,\lim_{\epsilon \downarrow0}\,\liminf_{r\downarrow0}
\por(\mu,x,r,\epsilon )~,
\end{equation}
and finally
\begin{equation}\label{eq@intropor}
\por(\mu)\,=\,\inf\{s~:~ \por(\mu,x)\le s \text{ for }\mu\text{-almost
all }x\in\real^n\}~.
\end{equation}
Note that when $x$ is not in the support of the measure, the
definition is not very interesting since for small enough $r$ the
measure of $\mu(B(x,r))$ is zero and then any ball in it is also empty.
\begin{@MyDefinition}The quantity $\por(\mu)$ is called the porosity
of the measure $\mu$.
\end{@MyDefinition}
\FIGUREWITHTEX{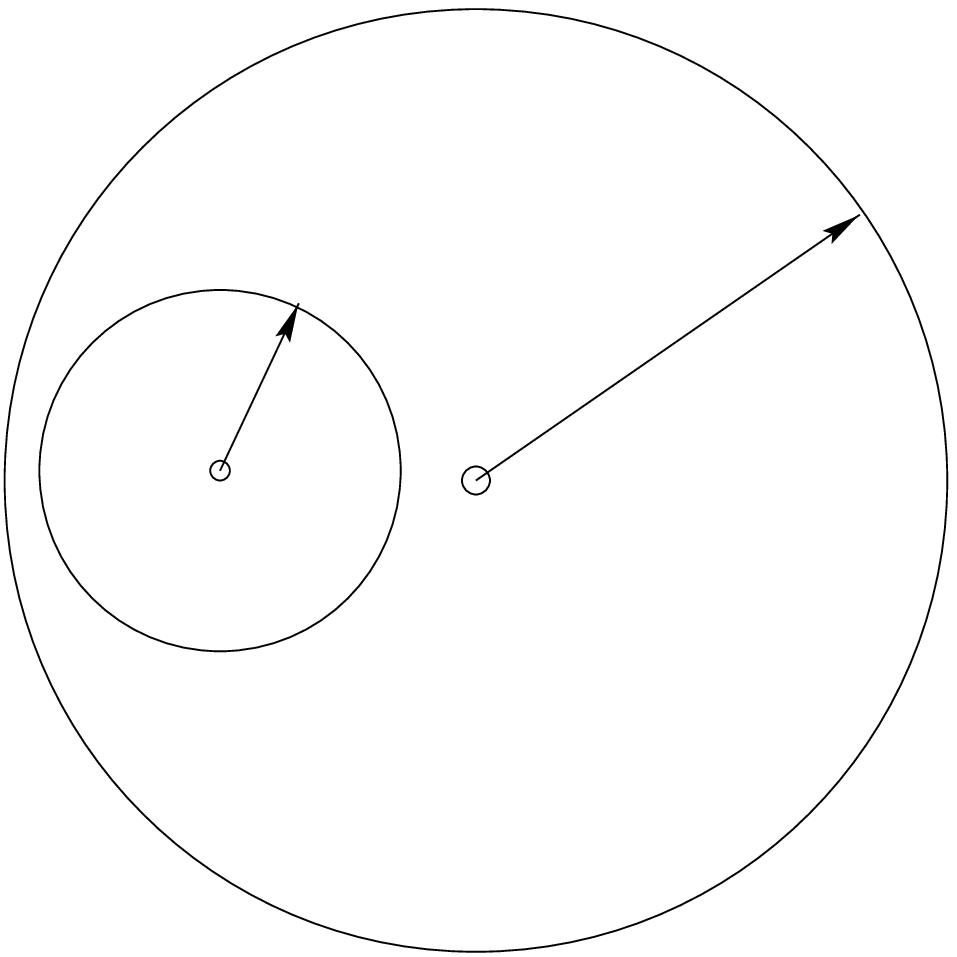}{fig1.tex}{100}{1}{0}{-0.9}{1}{The ball
$B(z,p r)$ inside $B(x,r)$.}

It is not difficult to see that the definition of the {\em porosity of a
set} (see Definition \ref{definition@defpor} below) amounts to using
Eq.\equ(pmx) with the 
limits taken in the 
opposite order, that is,
\be{pmx2}
\por(\spt(\mu),x)\,=\,\liminf_{r\downarrow0}\,\lim_{\epsilon \downarrow0}
\por(\mu,x,r,\epsilon )~.
\ee
Because the central point in $B(x,r)$ is in the support $\spt(\mu)$
and hence occupied when we compute $\por(\spt(\mu),x)$, one finds that
$\por(\spt(\mu))\le\frac12$. One can also show, using density
arguments, that
$\por(\mu)\le \frac12$. We also note that $\por$
is determined {\em first} with ``dust'' of relative weight $\epsilon $ and only
then $\epsilon$ is taken to 0. 

The two porosities we consider satisfy clearly
$\por(\spt(\mu))\le\por(\mu)$. In other words the porosity of a
measure is larger than that of the support of the measure, precisely
because the former neglects occasional dust.

\LIKEREMARK{Example 1}
Let $\delta_0$ be the Dirac measure at the origin, that is,
$\delta_0(A)=1$ if $0\in A$ and $\delta_0(A)=0$ if $0\notin A$. 
Let $\mu$ be the sum of $\delta_0$ and the Lebesgue measure ${\cal L}^n$
restricted to $B(0,1)$, that is, $\mu=C(\delta_0+{\cal L}^n|_{B(0,1)})$
where $C$ is the normalization constant. Clearly
$\por(\mu,0)=\frac 12$ and $\por(\mu,x)=0$ for all $x\ne0$ with $|x|<1$.
Thus $\por(\mu)=\frac 12$. However, $\por(\spt(\mu))=\por(B(0,1))=0$.

\LIKEREMARK{Example 2}We next want to argue that voids are accounted
for in a more 
reasonable way in the measure theoretic definition of porosity.
To illustrate this with a concrete example,
consider the celebrated middle third Cantor set which is
obtained by starting with the interval $[0,1]$ and taking out the open
interval $(\frac13,\frac23)$. Then each of the remaining two intervals
is divided into
three pieces and the middle one (of length $\frac19$) is discarded. Going on
recursively (and indefinitely) in this fashion, we get the middle
third Cantor set. Its 
dimension is $d\equiv \frac{\log2}{\log3}$. Clearly, this set $C$ has voids
and its porosity equals in fact $\frac14$.
We next set $C_x=\{ y~:~ y=x+z, z\in C\}$, in other words, $C_x$ is
the translate
of $C$ by $x$. Clearly each $C_x$ has again dimension $d$. From
the general theory of fractals\cite{F,M2} we get that any countable union of
such sets has still dimension $d$. In particular, we can enumerate the
rationals in $[0,1]$, for example calling them $x_j$, $j=1,2,\dots$ and
construct then the set
$$
D\,=\,\bigcup_{j=1}^\infty C_{x_i}~.
$$
From what we said before, and from the way we constructed it, the set
$D$ has dimension $d<1$ and {\em no voids}. Thus, we see that a {\em
set} can have small dimension and no voids.
The example we have just given does {\em not work} in the case of porosity
of measures and this is one of the reasons why the porosity of
measures is a more useful concept than that of sets.
However, we will construct regular fractals in Example 3 where the
porosity of the measure is arbitrarily small but the dimension is
always $\frac12$.

In fact, one shows easily, see\cite{EJJ},
that the following measure has porosity $\frac 12$.
We construct a measure on the set $D$ by giving successively lower
weight to the translates of $C$. Let $\mu $ be the usual measure
associated with the Cantor set $C$, \ie, the measure which gives equal
weight to all the pieces in the recursive construction. Then we define
$$
\nu\,=\,\sum _{i=1}^\infty 2^{-i} \mu_{x_i}~,
$$
where $\mu_{x}$ is the translate of the measure $\mu$ by $x$. Clearly,
the support of this measure is at least all of the interval $[0,1]$
(with some overhangs from the translation)
and has therefore no voids. But the porosity $\por(\nu)$, as defined in
Eq.\equ(intropor), is strictly positive. The reason for this is that if
we consider a void of the original set $C$ and look for a point in one
of the $C_{x_i}$ very
close to the boundary of this void, then we must take in general a
high index $i$ to find such a point. But then the associated
weight is smaller than $2^{-i}$,
and if $i$ is large enough, then this is smaller than any $\epsilon $
which was given in the definition of Eq.\equ(pmx). Therefore, the set
$C_{x_i}$ is not counted in this consideration, and the measure
will have the same porosity as $C$ itself.

Having described the definitions, we can now ask more precisely our question
about the relation between the porosity of the measure $\mu$ and the
packing dimension of the same measure. Our general aim is to show the
following
\begin{@MyConjecture}If the porosity of $\mu$ is large, then the
packing dimension of $\mu$ is smaller than the dimension $n$ of the
ambient space.
\end{@MyConjecture}

Our results will fall somewhat short of this conjecture.
In order to be able to formulate a positive result, we need the
following concept:
\begin{@MyDefinition}The probability measure $\mu$ on $\real^n$
satisfies the local doubling condition at $x$ if
\begin{equation}\label{eq@20}
\limsup_{r\downarrow0} {\m(x,2r)\over \m(x,r)}\,<\,\infty ~.
\end{equation}
It satisfies the (global) doubling condition if Eq.\equ(20) holds for
$\mu$-almost
every $x$.
\end{@MyDefinition}

The bound need not be uniform in those $x$.
Note that the doubling condition is
also implied by the stronger condition
\be{stronger}
0\,<\, a r^s \,\le\, \mu(B(x,r))\,\le\,b r^s \,<\,\infty ~.
\ee
In physics, it is generally assumed that the stronger condition
\equ(stronger) holds. 
See below for the relevance of these conditions in Nature.
Our main result is the following
\begin{@MyTheorem}\label{theorem@main}There is a function $\Delta_n$ defined for
$p\in(0,1/2)$ with values in $[0,1]$ and satisfying
$$
\lim_{p\to{1\over 2}} \Delta_n(p)\,=\,1~,
$$
such that
if a Borel probability measure $\mu$ on
$\real^n$ satisfies the global doubling condition
then
\begin{equation}\label{eq@ineq}
\dimH(\mu)\,\le\,\dimp(\mu)\,\le\, n-\Delta_n(\por(\mu))~.
\end{equation}
\end{@MyTheorem}
Here, $\dimH(\mu)$ is the Hausdorff dimension of the measure and
$\dimp(\mu)$ is its packing dimension. The inequality among those two is
obvious from their definition:
\begin{eqnarray*}
\underline
d(\mu,x)&\,=\,&\liminf_{r\downarrow0}{\log\mu(B(x,r))\over\log r}~,\\
\overline d(\mu,x)&\,=&\,\limsup_{r\downarrow0}{\log
\mu(B(x,r))\over\log r}~,\\
\dimH(\mu)&\,=\,&\sup\{s\ge0~:~
  \underline d(\mu,x)\ge s\text{ for }\mu
  \text{-almost all }x\in \real^n\}~,\\
\dimp(\mu)&\,=\,&\sup\{s\ge0~:~
  \overline d(\mu,x)\ge s\text{ for }\mu
  \text{-almost all }x\in\real^n\}~.
\end{eqnarray*}

There is an explicit lower bound for the function
$\Delta_n$ in\cite{S}: 
$$
\Delta_n(p)\,\ge\,\max\{1 - {c_n\over \log(\frac1{1-2p})},0\}~, 
$$
where $c_n>0$ is a constant depending only on $n$.
According to Theorem \ref{theorem@main} if the porosity of a measure $\mu$ which satisfies
the doubling condition is close to $\frac12$, then the packing 
dimension of $\mu$ is not much bigger than $n-1$. 

\REMARK For sets, a relation between porosity and dimension
has been established by Mattila\cite{M1} and Salli\cite{S} using the
following definition of porosity:
\begin{@MyDefinition}\label{definition@defpor}The porosity of a set $A\subset\real ^n$ at
a point $x\in\real ^n$ is defined by
$$
\por(A,x)\,=\,\liminf_{r\downarrow0}{\por(A,x,r)}~,
$$
where
$$
\por(A,x,r)=\sup\{ p\ge0:\,\text{ there is }z\in\real ^n \text{
such that } B(z,p r)\subset B(x,r)\setminus A\}.
$$
Here, $B(z,\alpha)$ is the closed ball with radius $\alpha$ and with center at
$z$.
The porosity of $A\subset\real ^n$ is
$$
\por(A)=\inf\{\por(A,x): x\in A\}~.
$$
\end{@MyDefinition}
We do not
know any example where
\begin{list}{}{}
\item{$\bullet$\ }The porosity of the measure is $\frac12$,
\item{$\bullet$\ }and the dimension of the measure is $n$ (in the ambient space
$\real^n$).
\end{list}

\noindent Of course, this would have to be a measure which violates
the doubling 
condition.

The doubling condition is a bound on the amplitudes of the
fluctuations of the integrated density of the measure. The role of
this condition in experiments in galaxy distributions is somewhat
obscure. 
But, for example, in\cite{MPST} the authors measured 2 periods of
density fluctuations (of about the same amplitude) and so in a very
weak sense, the doubling condition seems experimentally
satisfied. 
Another example is given in \fig(doublingp.ps) where the
results of measuring the doubling condition for certain catalogs of
galaxies are shown.
\FIGUREWITHTEX{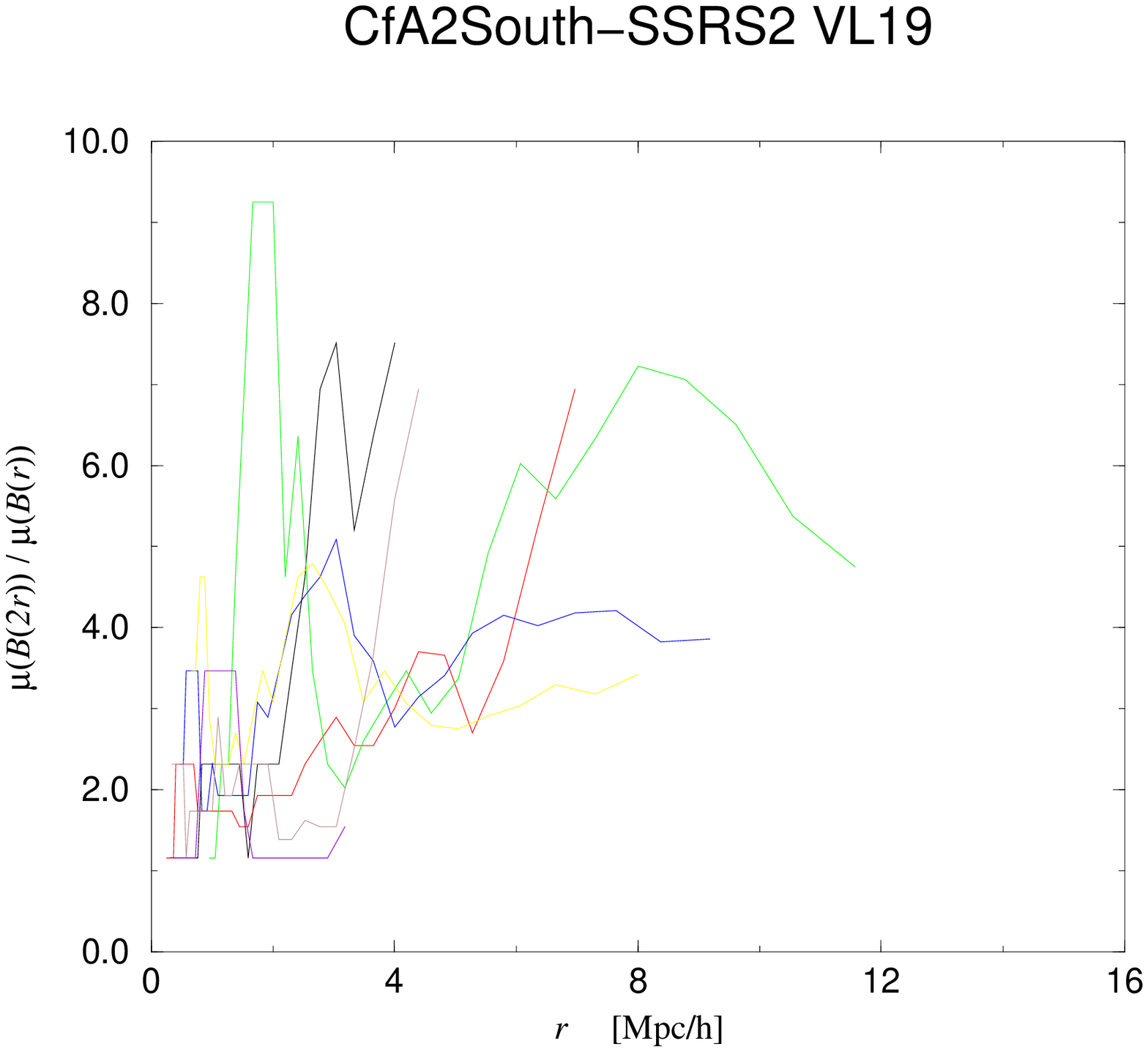}{null.tex}{1}{0.65}{0}{-0.65}{0}{Verification
of the doubling condition for several central points of two galaxy
catalogs. We thank M. Montuori and F. Sylos-Labini for these calculations.
The fluctuations seem quite bounded, except at short distances, where
there are problems because one runs out of data.}

For regular recursively constructed fractals the doubling condition
is always satisfied. Finally, we believe that fractals formed in
Nature by a physical law have not only the same dimension everywhere,
but also satisfy the doubling condition. The reason for this is that for
example an attractor looks everywhere similar because it is
created by a (smooth) physical law, which transports the structure of
the fractal around in space (making at most smooth coordinate changes
locally). This point of view has been advocated in\cite{ER}, and has
been rigorously verified for a few non-trivial examples of dynamical systems.

Another class of measurements takes the density
fluctuations of the mass itself as an indicator of the dimension of
the measure. This seems a mathematically inaccessible (and probably
wrong) criterion for dimension measurements. It might be that this
idea is a consequence of the regular oscillations one gets for regular
Cantor sets. The only case where a rigorous result is known is that of
integer dimension:

\begin{@MyTheorem}\label{theorem@Mar}(Marstrand) Let $s$ be a positive integer.
Suppose that there exists a Radon measure $\mu$ on $\real^n$ such that the
density
$$
\lim_{r\downarrow 0}{\mu(B(x,r))\over r^s}
$$
exists and is positive and finite in a set of positive $\mu$-measure. Then $s$
is an integer.
\end{@MyTheorem}

(For the proof see\cite{M2} Theorem 14.10.)

It is well-known that one cannot expect an inequality in the sense opposite
to the one stated in the theorem. That is, big voids imply small
dimension, but small
voids do not imply big dimension. This is illustrated by the
following example in $\real$, \ie, in one dimension.

\LIKEREMARK{Example 3}We will construct a sequence of measures
$\mu^{(n)}$, all of dimension $\dimp(\mu^{(n)})={1\over 2}$ in $\real$ with
porosity
$\por(\mu^{(n)})\le {1\over n}$. The set $A^{(n)}$ is a Cantor set obtained
recursively as
follows: Divide the interval $[0,1]$ into $n^2$ equal subintervals
and select $n$ of these subintervals, namely the $1^{\text {st}}$,
$n+1^{\text{st}}$, and so on.
The measure at this level of the construction
is obtained by giving the same weight ${1\over n}$ to each
subinterval of $A^{(n)}$. In other words, the unit measure is
uniformly distributed on the $n$ intervals constructed so far.
\FIGUREWITHTEX{lacundivp.ps}{lacundiv.tex}{1}{0.54}{0}{-0.65}{1}{The
oscillations of $\por(r)$ as a function of $r$ (on a logarithmic scale)
for the sets $A^{(n)}$, with $n=3,\dots,6$.}

Now repeat inductively the procedure for
each of the $n$ intervals, dividing it into $n^2$ equal pieces and
selecting each $n^{\text{th}}$ among them. Give each of these
intervals weight $\frac1{n^2}$.
\FIGUREWITHTEX{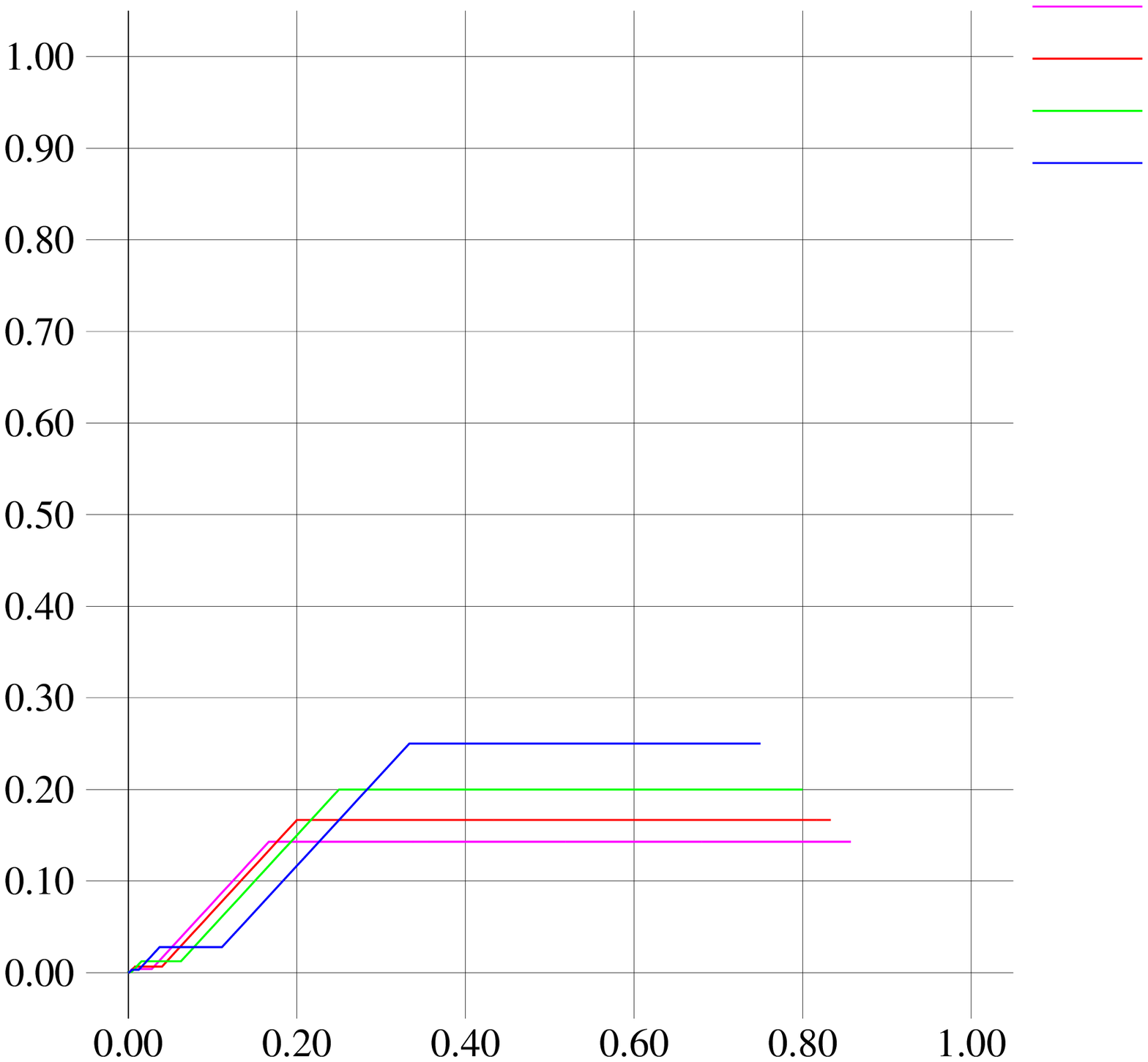}{lacunnodiv.tex}{1}{0.54}{0}{-0.65}{1}{The
largest void $\rho$ as a function of $r$, measured from the left most
point in $A^{(n)}$
for $n=3,\dots,6$.}

Continuing indefinitely in this fashion, one obtains the Cantor set
$A^{(n)}$ and
the measure $\mu^{(n)}$ on it.
The dimension of this measure is $\dimp(\mu^{(n)})= \log( n) / \log
(n^2)=\frac12$, and it is not difficult to check that the porosity is
less than
$\frac1n$. (Since the gaps become smaller for larger $n$, see\cite{EJJ}.)
In this case, it is easy to compute numerically the porosity of the
sets $A^{(n)}$ when $n$ is not too large. In \fig(lacundivp.ps) we 
show the quotient $\por(\mu,x,r,\epsilon =0)$ as a function of $r$
when $x$
is the leftmost point of $A^{(n)}$. Similar, but more irregular
pictures are obtained when one chooses another point $x$.
One can understand the origin of the oscillations by looking at
\fig(lacunnodivp.ps), where we plot the radius $\rho$ of the
largest empty interval as a function of $r$.
We see that $\rho $ grows linearly, until a point of the Cantor set is
hit, and then it stays constant until a bigger void is found.
Then \fig(lacundivp.ps) is obtained by dividing the values obtained
in \fig(lacunnodivp.ps) by $r$.

The proof of our main result is based on comparing the porosity of a
measure with the porosity of subsets with positive measure. For this,
we use the quantity $\beta (\mu)$ introduced in\cite{MM}:
$$
\beta(\mu)\,=\,\sup\{\por(A): A\text { is a Borel set with }\mu(A)>0\}~.
$$
The inequality $\beta (\mu)\le\por(\mu)$ holds for any Borel
probability measure $\mu$, but the converse inequality does not need to be
true. We show that it holds when the measure $\mu$ satisfies the
doubling condition. We have shown in\cite{EJJ} that the doubling
condition implies
 $\beta
(\mu)=\por(\mu)$. Together with the results of\cite{S} it implies
our main bound  on the dimension (see Theorem \ref{theorem@main}). 
We also showed that there
are measures violating the doubling condition for which  $\beta
(\mu)\ne\por(\mu)$. In the case of measures on the line $\real $, {\it
i.e.,} in 1 dimension, we
also show that a somewhat weaker condition than the doubling condition
implies  Theorem \ref{theorem@main}.

\end{document}